\documentclass[aps,prl,twocolumn,groupedaddress,amssymb]{revtex4}
\usepackage{graphicx,latexsym}

\begin{document}
\def\be{\begin{equation}}
\def\ee{\end{equation}}
\def\ba{\begin{eqnarray}}
\def\ea{\end{eqnarray}}

\title{Ensemble Averaged Conductance Fluctuations in Anderson Localized Systems}

\author{M. Hilke}
\email{hilke@physics.mcgill.ca}
\affiliation{Dept. of Physics, McGill University, Montr\'eal, Qu\'ebec, H3A 2T8}
\affiliation{Institute for Theoretical Physics, TU Berlin, 10623 Berlin, Germany}



\begin{abstract}
We demonstrate the presence of energy dependent fluctuations in the localization length, which depend on the disorder distribution. These fluctuations lead to Ensemble Averaged Conductance Fluctuations (EACF) and are enhanced by large disorder. For the binary distribution the fluctuations are strongly enhanced in comparison to the Gaussian and uniform distributions. These results have important implications on ensemble averaged quantities, such as the transmission through quantum wires, where fluctuations can subsist to very high temperatures. For the non-fluctuating part of
the localization length in one dimension we obtained an improved analytical expression valid for all disorder strengths by averaging the probability density.
\end{abstract}

\maketitle
In quantum coherent conductors, disorder induces dramatic effects on the conductance. For instance, if the localization length, $L_c$, which is the decay length of the wavefunction, exceeds the system size $L$, the conductance vanishes exponentially (Anderson localization) \cite{Anderson58}. On the other hand, if the system size is smaller than $L_c$, then the conductance exhibits fluctuations in energy which take on universal amplitudes and are referred to as Universal conductance fluctuations (UCF) \cite{Lee85}. In one dimension, the effect of disorder is particularly dramatic because the presence of any amount of uncorrelated disorder will lead to a finite localization length at all energies. The localization length, which is an ensemble averaged quantity, has typically no fluctuations as a function of energy. This is in stark contrast to the conductance for a single disorder configuration, where the fluctuations are maximized
due to UCF.

In many quantum wires, such as carbon nanotubes, the conductance can be measured as a function of the gate voltage, effectively changing the Fermi energy of the carriers. As a function of
energy, the conductance often shows strong fluctuations associated with disorder. The amplitude of these fluctuations decays with temperature \cite{Stojetz04,Man05}. This can be understood
in terms of ensemble averaging, since an increase in temperature leads to a decrease of the quantum coherence length $l_\phi$ due to increased
inelastic scattering. This in turn leads to the ensemble averaging over the quantum size $l_\phi$ averaged $L/l_\phi$ times for a sample of size $L$. it is therefore surprising that experimentally, the conductance fluctuations of carbon nanotubes remain substantial even at high temperatures
(equivalent to ensemble averaging) \cite{Stojetz04,Man05}.

\begin{figure}[!htb]
    \begin{center}
    \vspace*{0cm}
    \includegraphics[scale=0.8]{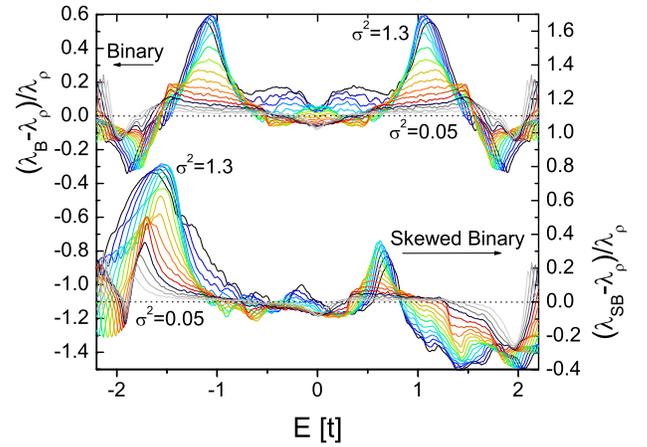}
    \vspace*{-0cm} \caption{The relative deviation of the Lyapounov exponent for the binary and skewed binary distribution. The skewed binary distribution function of average zero is given by $P(V)=(1-\alpha)\delta(V-\alpha W)+\alpha\delta(V+(1-\alpha)W)$, where we have $\alpha=1/2$ for the binary case ($\lambda_B$) and we chose $\alpha=1/3$ for the skewed case ($\lambda_{SB}$). The variance $\sigma^2$ is varied between .05 and 1.3 and depends on $W$. $\lambda_\rho$ is the Lyapounov exponent obtained from the average probability density.}
    \label{Distribution}
    \end{center}
\end{figure}

\begin{figure}[!htb]
    \begin{center}
    \vspace*{0cm}
    \includegraphics[scale=0.8]{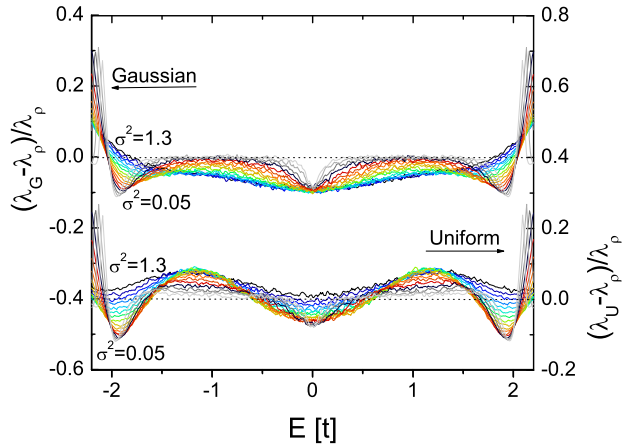}
    \vspace*{-0cm} \caption{The relative deviation of the Lyapounov exponent for the Gaussian distribution ($\lambda_G$) of zero average and of standard deviation $\sigma$ and the uniform distribution ($\lambda_U$ with $-W/2<V<W/2$ and $\sigma^2=W^2/12$). $\sigma^2$ is varied between 0.05 and 1.3. }
    \label{Distribution2}
    \end{center}
\end{figure}

In order to better understand possible fluctuations even after ensemble averaging, we studied in detail the energy dependence of the ensemble averaged quantity $L_c$. Quite strikingly, we observe fluctuations of $L_c$ as a function of energy, which depend on the distribution of the disorder potential. This is illustrated in Figures \ref{Distribution} and \ref{Distribution2}, where we show the relative variation of the inverse
localization length (Lyapounov exponent, $\lambda$) as a function of energy for different disorder distributions (Gaussian, uniform and binary). For the calculations we used the
one-dimensional Anderson model, described in more detail below.

The variations (or fluctuations) of the Lyapounov exponent are quite substantial for the binary distribution and reach up to 80\%. The fluctuations are shown with respect to the Lyapounov exponent ($\lambda_\rho$) obtained from the average probability density, valid for all disorder strengths and derived below. The symmetry of the fluctuations depends on the symmetry of the distribution. Indeed, for $\alpha=1/2$ the distribution is symmetric
in energy,
which leads to symmetric fluctuations as seen in Figure \ref{Distribution} in stark contrast to the skewed case $\alpha\neq 1/2$, where the distribution is asymmetric.

In Figure \ref{Distribution2} we show the relative deviations of the Lyapouvov exponent for the Gaussian and uniform distributions. Here the fluctuations are strongly suppressed and the deviations are of the order of 10\% and are very smooth when compared to the binary distributions for the same standard deviation range. However, the observed strong fluctuations for the binary distributions, lead us to make a comparison with mesoscopic conductance fluctuations.

In general, conductance fluctuations are expressed as $\langle \delta G^2\rangle=\langle (G-\langle G \rangle)^2\rangle$, where $\langle \cdot \rangle$ denotes the ensemble average, i.e., the average over a given disorder distribution. The standard universal conductance fluctuations (UCF) result gives $\sqrt{\langle \delta G^2\rangle}\simeq 0.73e^2/h$ \cite{Lee85} for a quasi 1D system. In the localized regime these fluctuations are suppressed. In order to characterize the fluctuations in the Lyabounov exponent seen in Figure \ref{Distribution} and \ref{Distribution2}, we define
\be
\delta \lambda^2=\frac{1}{E_{max}-E_{min}}\int_{E_{min}}^{E_{max}}[\lambda(E)-\lambda_\rho(E)]^2\mbox{dE},
\ee
where $\lambda_\rho$ is the Lyapounov exponent obtained from the average probability density with the same standard deviation. The bandwidth $(E_{max}-E_{min})$ is fixed to the non-disordered one.

\begin{figure}[!htb]
    \begin{center}
    \vspace*{0cm}
    \includegraphics[scale=0.8]{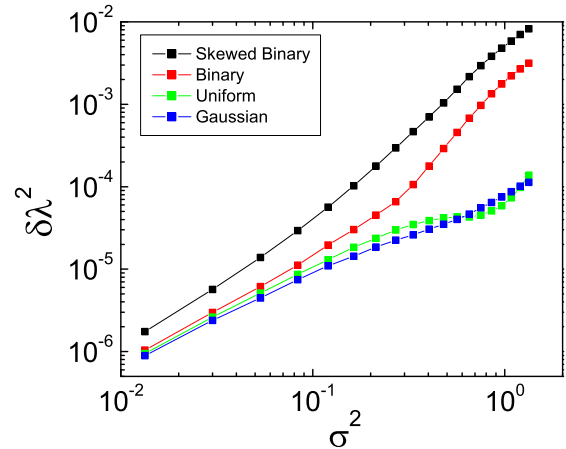}
    \vspace*{-0cm} \caption{$\delta\lambda^2$ is shown for different distributions as a function of the disorder strength, which is characterized
    by the variance $\sigma^2$. For the skewed binary distribution we used $\alpha=1/3$ (the same as in Figure \ref{Distribution}). These fluctuations represent ensemble averaged conductance fluctuations (EACF).}
    \label{ECF}
    \end{center}
\end{figure}

For a strongly localized system of size $L$, the ensemble averaged conductance is given by $\langle G \rangle \sim\exp^{-2\lambda L}$. Hence, the fluctuations in $\lambda$ lead to fluctuations in the ensemble-averaged conductance $\delta\langle G\rangle\sim 2\langle G\rangle L\delta\lambda$, which we coin ensemble averaged conductance fluctuations (EACF). We studied EACF for different disorder strengths. The results are shown in Figure \ref{ECF} for different distributions.

We now describe the derivation of the Lyapounov exponent ($\lambda_\rho$) for the average probability density, which is used in the calculations
of the relative fluctuations shown in Figure \ref{ECF}. This is a novel analytical approach in order to obtain an expression for $\lambda(E)$ valid for all disorder strengths.

The one-dimensional Anderson model \cite{Anderson58} is a tight binding equation with random on-site potentials $V_n$ and given by
\be t\Psi_{n+1}+t\Psi_{n-1}=(E-V_{n})\Psi_{n}. \label{TB1}\ee
Here $t$ is the hopping term, which we set to one. In this model, an alloy of two elements would be described with $V_n$ taken from
a binary distribution like in Figure \ref{Distribution}. In contrast, if the disorder is due to the surface roughness of the substrate, like in the case of a carbon nanotube on a silicon oxide
substrate, $V_n$ would be given by a more continuous distribution, like the Gaussian or uniform one.

In order to obtain $\lambda(E)$ the main idea is to obtain and iterative equation for the probability density and then to average it. Hence, assuming real potentials $V_n$ it is possible to rewrite (\ref{TB1}) as
\begin{widetext}
\ba
\rho_{n+1} & = & \left[(E-V_{n})^2-\frac{E-V_{n}}{\{E-V_{n-1}\}}\right]\rho_{n}+\left[1-(E-V_{n})(E-V_{n-1})\right]\rho_{n-1}
+\left[\frac{E-V_{n}}{E-V_{n-1}}\right]\rho_{n-2}\label{TBrho1}\\
 & = & \left[(E-V_{n})^2-1]\right]\rho_{n}+\left[1-2(E-V_n)(E-V_{n-1})+(E-V_{n-1})^2+\frac{V_n-V_{n-1}}{\{E-V_{n-2}\}}\right]\rho_{n-1} \label{TBrho2}
\\ & & +\left[(E-V_{n})(E-V_{n-2})+1-(E-V_{n-1})(E-V_{n-2})\right]\rho_{n-2}+\left[\frac{V_{n-1}-V_{n}}{E-V_{n-2}}\right]\rho_{n-3}\nonumber,
\ea
\end{widetext}
where $\rho_n=\psi_n\psi_n^*$ is the probability density. Equation (\ref{TBrho2}) is obtained by using an additional iteration and will be important when considering the average.
Interestingly, this exact (before averaging) iterative expression for the probability density depends explicitly on neighboring potentials, which illustrates the importance of the assumption of an uncorrelated disorder potential when taking the disorder average. Indeed, the presence of local correlations can lead to an infinite localization length in one and two dimensional
disordered systems \cite{corr}. In our method of considering the probability density, $\rho_n$ instead of $\psi_n$, phase correlations are not averaged out when the disorder average is performed.

The average of (\ref{TBrho2}) can be taken trivially by assuming uncorrelated disorder with $\langle V_nV_m\rangle=\sigma^2\delta_{n,m}$ and noting that $\langle \rho_i V_j \rangle=\langle \rho_i\rangle \langle V_j \rangle$ for $i\geq j$, since $\rho_n$ does not depend on $V_n$ when $\rho_n$ is obtained iteratively using initial conditions for $n=0,1,2,3$. The term in curly brackets of $\rho_n$ in (\ref{TBrho1}) is the only product which cannot be separated from $\rho_n$ when averaged, which is the reason we had to iterate this equation one more time in order to obtain equation (\ref{TBrho2}), where the coefficient in front of the curly bracket term now averages to zero.

For a potential of average zero we thus obtain the following iterative solution for the average probability density:

\be \langle\rho_{n+1}\rangle=(E^2+\sigma^2-1)\langle\rho_n\rangle+(1-E^2+\sigma^2)\langle\rho_{n-1}\rangle+
\langle\rho_{n-2}\rangle\label{rhoAV2}.\ee

\begin{figure}[!htb]
    \begin{center}
    \vspace*{0cm}
    \includegraphics[scale=0.8]{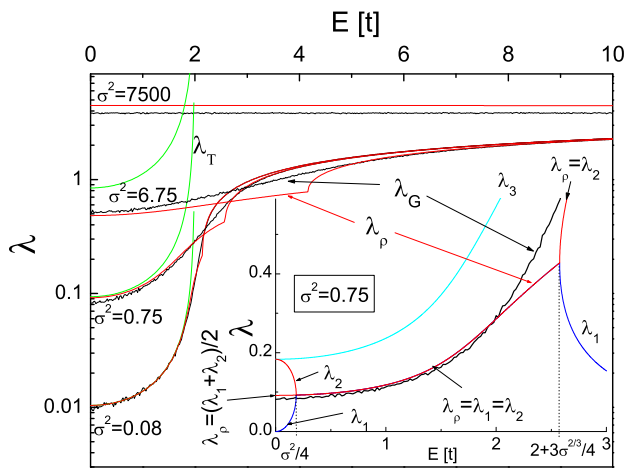}
    \vspace*{-0cm} \caption{Lyapounov exponents ($\lambda_G$, $\lambda_\rho$ and $\lambda_T$) as a function of energy for different disorder strengths. In the inset $\lambda_1$, $\lambda_2$, and $\lambda_3$ are shown together with $\lambda_G$ and $\lambda_\rho$.}
    \label{HighV}
    \end{center}
\end{figure}

The leading dependence of $\langle \rho_n\rangle$ can now be extracted by evaluating the eigenvalues $\{\xi_1,\xi_2,\xi_3\}$ of the characteristic transfer matrix determined by (\ref{rhoAV2}). The corresponding Lyapounov exponents are
\be \lambda_{i}=\frac{1}{2}|\log(|\xi_i|)|, \label{minmax}
\ee
where the factor $1/2$ comes from the definition of the Lyapounov exponent in terms of the probability density instead of the wavefunction. For $\sigma=0$ all three Lyapounov exponents are zero, which is characteristic of the plane wave solution for an energy within the band $-2\leq E\leq 2$. Defining $\lambda_1\leq \lambda_2\leq \lambda_3$ we have for $\sigma >0$ that $\lambda_1>0$, which implies that all states are localized. An exception occurs at $E=0$, where $\lambda_1=0$ and $\lambda_2>0$. However, the solution corresponding to $\lambda_1$ violates the positivity requirement of $\rho_n$. Hence, for energies close to $E=0$ we arbitrarily define $\lambda_\rho$ as the average of the two smallest Lyapouov exponents, instead of just the smallest exponent, i.e., $\lambda_\rho\equiv(\lambda_1+\lambda_2)/2$. Away from the band center, corresponding to
$|E|\gtrsim \sigma^2/4$, we have $\lambda_{\rho}\equiv\lambda_1=\lambda_2$. Here the smallest Lyapounov exponent $\lambda_1$ correctly describes the dependence. In the limit of small disorder ($\sigma\ll 1$) this leads to $\lambda_{\rho}\simeq (1/2)\sigma^2/(4-E^2)=\lambda_T(E)$, where $\lambda_T$ corresponds to the standard result by Thouless \cite{Thouless79}, which was obtained by averaging the Green's function to second order.
For $|E| \gtrsim 2+3\sigma^{2/3}/4$ ($\sigma\ll 1$) we have
$\lambda_1\neq\lambda_2$ and the solution corresponding to $\lambda_1$ violates the positivity requirement of $\rho_n$, which is why we define $\lambda_\rho\equiv\lambda_2$ in that range, which is outside of the non-disordered band. The behavior of the Lyapounov exponents corresponding
to the three eigenvalues are shown in the inset of Figure \ref{HighV} and are similar to the ones studied for continuous disordered
systems \cite{Erdos89}.

For the special energy at the band center, the correction for the small disorder expansion \cite{Kappus81,Derrida84} is given by $\lambda^{K-W}(E=0)\simeq 0.91\lambda^T(0)$, where $\lambda^{K-W}$ is the correction due to the Kappus-Wegner anomaly. This anomaly at the band center can also be seen in the context of our approach using the probability density, since expressions (\ref{TBrho1} and \ref{TBrho2}) are ill defined when taking the average at $E=0$. At the band edge the first order correction to the Thouless result was obtained recently by using a
classical two-dimensional Hamiltonian map \cite{Tessieri00}.

The anomalies at the band center and at the band edges can be easily identified in Figure \ref{Distribution2} for the Gaussian distribution, where they appear as bumps for small disorder. Most previous analytical approaches are based on small disorder expansions and cannot be applied to high disorder. This is in contrast to our probability density approach, where we didn't assume small disorder, and we indeed obtain the correct large disorder limit $\lambda_{\rho}\equiv\lambda_2\simeq\lambda_{3}\simeq\log(\sigma)$ with a Lyapounov exponent independent of $E$ \cite{Thouless72}.

Numerically, the Lyapounov exponent was obtained by evaluating the eigenvalues of the product of transfer matrices obtained from equation (\ref{TB1}),
\be \{\xi^N_i\}=\mbox{eig}\left[\prod_{n=1}^N\left(\begin{array}{cc}
E-V_n & -1 \\1 & 0 \end{array}\right)\right]
\ee
and then taking the self-averaging limit where we used $N\gg 10^6$.
\be \lambda=\lim_{N\rightarrow\infty}\frac{1}{N}\log(\max\{|\xi^N_i|\})=-\lim_{N\rightarrow\infty}\frac{1}{N}\log(\min\{|\xi^N_i|\}).
\ee

The quality of our expression for $\lambda_\rho$ can be seen in Figure \ref{Distribution2}, where we show the relative deviation to the numerically obtained Lyapounov exponents $\lambda_{G,U}$. The deviations are of the same order as the dependence on the distribution function (Gaussian versus uniform). This shows that any further improvement of the analytical expression for $\lambda(E)$ needs to depend on the distribution function explicitly (for example by including higher moments). For the special case of the Cauchy distribution, where the
second moment does not exist, an exact analytical expression for the Lyapounov exponent can be found \cite{Ishii}.

We also compared our expression $\lambda_\rho$ in Figure \ref{HighV} for a large range of disorder strengths and find that the fit is less than a factor two off for all energies and disorder strengths, in contrast to the Thouless expression ($\lambda_T$) also shown, which deviates substantially at large disorder.

The ensemble averaged fluctuations, which is the main result presented here, were obtained for one of the simplest models showing Anderson localization, namely the random one-dimensional tight binding equation. However, Anderson localization is very general and can be observed,
among others, in the propagation of light in disordered media, in phonon and plasmon modes, in quantum chaotic systems \cite{phonon}, in Bose Einstein condensates \cite{BEC}, and even in neutron propagation \cite{neutron}. We therefore expect that similar EACF also exist in these systems, since the equations describing these
are very similar. For dimensions higher than one, only very few analytical expressions exist, including a recent result valid for small disorder, which has been derived for quasi one-dimensional systems \cite{Baldes04} and that can be expressed as a convolution of the one-dimensional case. We therefore expect that at higher disorder strengths, distribution dependent fluctuations will also appear in quasi one-dimensional systems, which are relevant for the experiments on quantum wires. More striking examples might be systems composed of alloys, since the potentials of a two component alloy would be described by a binary disorder distribution, which shows the strongest EACF. Indeed, in InGaAs alloys, conductance fluctuations in a quasi
one-dimensional geometry were observed at high temperatures \cite{Hackens02}. However, the study of EACF in higher dimensions is beyond the scope of this work.

\end{document}